\documentclass[9pt,twocolumn,twoside]{opticajnl}
\journal{opticajournal} % use for journal or Optica Open submissions

\setboolean{shortarticle}{true}
\usepackage{lineno}
\title{Zernike Mode Sorting with Vortex Phase Filters: Perfect Coronagraphs and Ideal Wavefront Sensors}

\author[1, *]{Jacob Trzaska}
\author[1, 2]{Amit Ashok}

\affil[1]{Wyant College of Optical Sciences, University of Arizona, Tucson, AZ 85721, USA}
\affil[2]{Department of Electrical and Computer Engineering, University of Arizona, Tucson, AZ 85721, USA}
\affil[*]{jtrzaska@arizona.edu}

\begin{abstract}
Spatial mode sorting has come to prominence as an optical processing modality capable of saturating fundamental limits to numerous sensing tasks including wavefront sensing, coronagraphy, and superresolution imaging. But despite their promising theoretical advantages, contemporary mode sorters often feature large crosstalk, high loss, or sort modes that are poorly adapted to conventional imaging systems (e.g., Hermite- and Laguerre-Gauss). Here, we introduce an alternative architecture that sorts spatial modes natural to circularly symmetric apertures: Zernike polynomials. Using conventional optics hardware and even-order vortex phase plates, we show how to assemble a series of vortex phase filters that can in principle separate the various Zernike polynomials losslessly and without crosstalk. This idea is demonstrated via application to wavefront sensing and coronagraphy, where we propose an optical system that saturates the quantum sensitivity limits to both tasks. We expect our work to prove useful for high-contrast imaging of extrasolar planets, improving both wavefront control and coronagraph performance.
\end{abstract}

\setboolean{displaycopyright}{false} % Do not include copyright or licensing information in submission.

\begin{document}
\maketitle
\section{Introduction}\label{sec:intro}
Spatial mode sorting is an optical processing protocol that demultiplexes an incident field's constituent spatial modes. This operation generalizes that of a paraxial lens, which sorts spatial frequencies into its focal plane. At focus we can then directly access an object's Fourier spectrum and either modulate the field via filtering or simply form an image. Likewise, we can devise spatial mode sorters capable of demutiplexing any collection of orthogonal functions, greatly expanding the set of optically-accessible mathematical operations and electronic measurements. This is perhaps the most utilitarian reason for optical demultiplexing, as a slew of quantum information analyses over the past decade have found that sorting the optical field into specially engineered spatial modes can offer superior sensitivity relative to traditional focal plane arrays \cite{tsang_quantum_2016, Rehacek:17, ronan, PhysRevA.99.022116, grace-thesis, PhysRevA.107.022409, PhysRevA.107.032427}. These advantages were perhaps best illustrated by Tsang et al. \cite{tsang_quantum_2016, Rehacek:17, ronan} where those authors suggested that measuring photons in aperture-adapted spatial modes allows resolving point sources well below the Rayleigh diffraction limit.

This pioneering idea has since been applied extensively to astronomical problems \cite{2023aoel.confE..79H, PhysRevResearch.6.013212, Trzaska:24, deshler2025quantumlimitsexoplanetdetection, deshler_experimental_2025}, with follow-on studies suggesting that state-of-the-art astronomical instruments fail to reach their fundamental performance ceilings. Amongst these systems are wavefront sensors (WFS) \cite{guyon2018AnnualReviews} and coronagraphs \cite{kenworthyHaffert2025}, the two apparatus most critical for directly imaging extrasolar planets. Indeed, quantum information theory tells us that a direct imaging WFS or coronagraph must extract the piston mode and either phase shift it or null it, respectively, for maximum performance \cite{guyon_limits_2005, chambouleyron_coronagraph-based_2024, guyon_theoretical_2006, deshler_experimental_2025}. Currently, no WFS or coronagraph sieves piston perfectly \cite{guyon_high_2010, deshler2025quantumlimitsexoplanetdetection}. 

A survey of existing mode sorting technology shows that this capability remains out of reach. For example, photonic integrated circuits (PICs) are a promising hardware platform for reconfigurable mode sorter. Paranal Obesrvatory has already employ PICs for beam combination at the Very Large Telescope Interferometer \cite{gravity_collaboration_first_2017} and Belikov et. al. \cite{astroPIC} has demonstrated an integrated coronagraph in laboratory that achieves 10\textsuperscript{-7} contrast. However, these devices are currently lossy and sort a limited number of spatial modes. Another modality photonic lanterns (PLs) \cite{Birks:15, vievard2024, Lin:25} furnishes excellent mode counts (> 1000) and are already used for both wavefront sensing and spectroscopic tasks on ground-based telescopes. But PLs are also lossy and their mode bases are difficult to control and characterize, resulting in substantial crosstalk and loss of sensitivity. We finally have multiplane light converters (MPLCs), which use a series of phase masks coupled through free space propagation to achieve arbitrary spatial mode transformations \cite{labroille_efficient_2014, fontaine_design_2017}. MPLC has already demonstrated its potential in a recent coronagraphy experiment \cite{deshler_experimental_2025}, where the authors successfully localized a dim point source (1000:1 contrast) at sub-Rayleigh scales. Like a PIC, MPLC offers reconfigurability when implemented with spatial light modulators (LCoS or deformable mirror). Unfortunately the size, cost, and computing requirements for training an ultra-low crosstalk MPLC scale poorly with mode count. These difficulties motivate the need for a new strategy.

This work presents a novel mode sorting modality based on the principles underlying vortex coronagraphy \cite{Foo:05}. In particular, we develop a general theory of \emph{vortex phase filtering} by adopting the even-order vortex phase mask as a primitive optical component. In conjunction with lenses, beasmplitters, and phase shifters, we show how to an assemble an optical system than can in principle sort the Zernike polynomials losslessly and without crosstalk. We then demonstrate how cascading two of these filters is enough to isolate Zernike piston, the resource extraction necessary for saturating the fundamental limits to single-conjugate wavefront sensing and exoplanet detection around unresolved stars. Extensions to arbitrary apertures and arbitrary mode bases are presented thereafter.

\section{Zernike Mode Sorting}\label{sec:sorting}
\begin{figure}
    \centering
    \includegraphics[width=\linewidth]{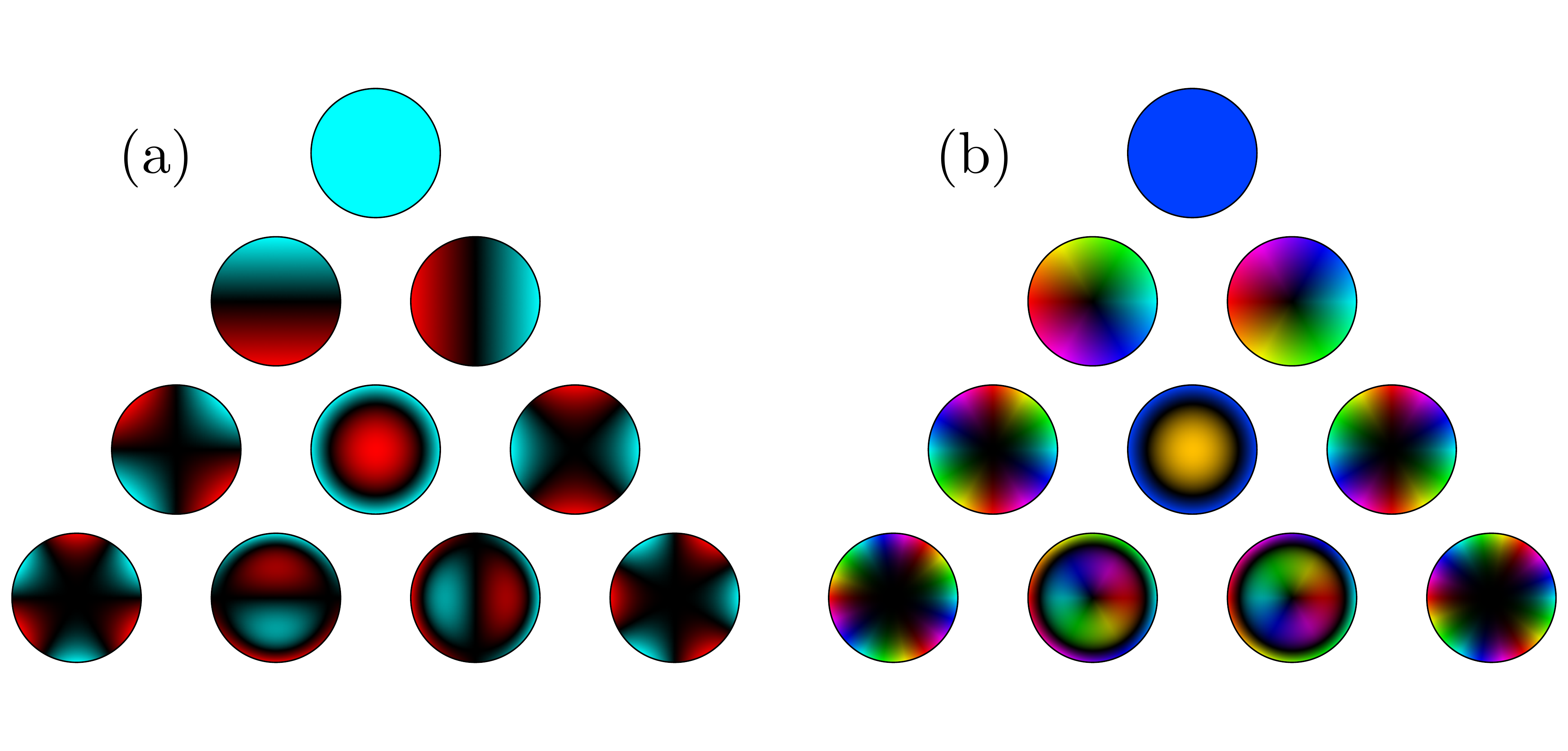}
    \caption{(a) First ten real- and (b) complex-valued Zernike polynomials. In both plots value represents mode amplitude and hue the phase. Real Zernikes are ubiquitous in optics, but here complex Zernikes prove more valuable because of their definite OAM, a feature vital for developing a theoretically perfect mode sorter.}
    \label{fig:zernikes}
\end{figure}

Zernike polynomials \cite{zernike_diffraction_1934, Bhatia_Wolf_1954, Niu_2022} are the natural basis for functions on circular apertures. They come in two varieties, one real-valued (Fig. \ref{fig:zernikes}(a)) and the other complex-valued (Fig. \ref{fig:zernikes}(b)). Real-valued modes $\mathcal{Z}_n^m$ dominate metrological and imaging applications due to their describing classical aberrations such as tip, tilt, and defocus. Complex Zernikes $Z_n^{\pm|m|} = \mathcal{Z}_n^{|m|} \pm i\mathcal{Z}_n^{-|m|}$ are seldom employed but have the useful feature that they carry integer orbital angular momentum (OAM). In other words, their phase evolves linearly in azimuth, corresponding to a type of vortex wavefront evident in Fig. \ref{fig:zernikes}(b). Incredibly, this definite OAM greatly facilitates their spatial sorting.

\begin{figure}
    \centering
    \includegraphics[width=\linewidth]{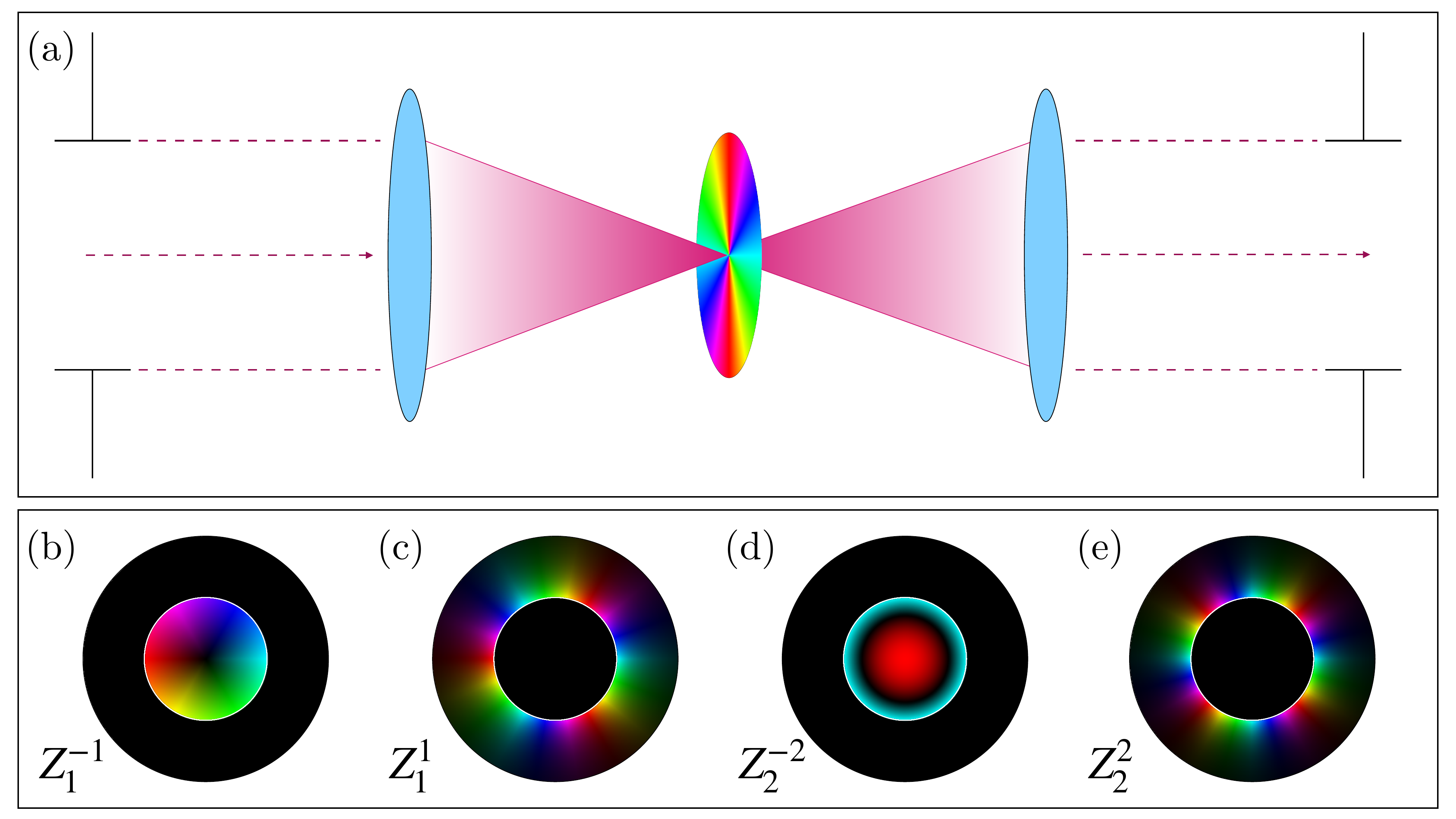}
    \caption{(a) Optical layout of a VPF, shown here as charge 2. Light from the pupil is brought to focus and phase shifted using a vortex phase mask. A second lens inverts the imaging, relaying the field to a conjugate pupil plane. (b) - (e) Select complex Zernike polynomials (labels bottom left) after propagating through a VPF\textsubscript{2}. Pupil edges are marked by the white lines. Maximum radial distance shown is one diameter. Notice that modes $Z_n^m$ having $m \geq n$ are removed from the pupil while modes with $m < n$ have been simply transformed into $Z_{n}^{m+2}$. This type of nulling was first observed for piston by Foo et al. \cite{Foo:05} but manifests for any polynomial shifted to an invalid Zernike index.}
    \label{fig:vpf}
\end{figure}

To demultiplex the complex Zernike polynomials we propose using what we term a vortex phase filter of charge $l$, or a VPF\textsubscript{$l$}. Fig. \ref{fig:vpf}(a) shows an optical drawing of the device. It is simply a pupil relay built from two lenses (or mirrors) with a vortex phase plate situated at the lens' common focal point. In this work the  vortex charge is always even, i.e., $l=2k$. When a stop is placed at the conjugate plane the VPF is recognized as a vortex coronagraph (VC) \cite{Foo:05, mawet_annular_2005} or as a bivortex WFS (bvWFS) \cite{chambouleyron_coronagraph-based_2024}. However, it is presently more useful to vacate the pupil plane and better understand the action of a VPF on the complex Zernike basis. Indeed, let us propagate these modes through a VPF\textsubscript{$l$} and examine their spatial structure out the output plane (Appendix \ref{app:sec:vortex-math}). What we find is a remarkable mathematical identity \begin{equation}
    \label{eq:propagated-zernikes}
    VPF_l\left(Z_n^m\right) = \begin{cases}
        Z_n^{m+l}(r, \theta), & |m+l|\leq n, 2r<D\\
        0, & |m+l| > n, 2r < D
    \end{cases}.
\end{equation}That is, modes possessing OAM $|m+l|>n$ perfectly destructively interfere over the entire pupil! Fig. \ref{fig:vpf}(b)-(e) illustrates this effect for several low-order Zernike polynomials, showing that ejected modes exhibit rapidly decreasing irradiance outside the aperture's support. Foo et al. \cite{Foo:05} and Mawet et al. \cite{mawet_annular_2005} had first observed this behavior with piston, but vortex filtering is verily more versatile.

\begin{figure}
    \centering
    \includegraphics[width=\linewidth]{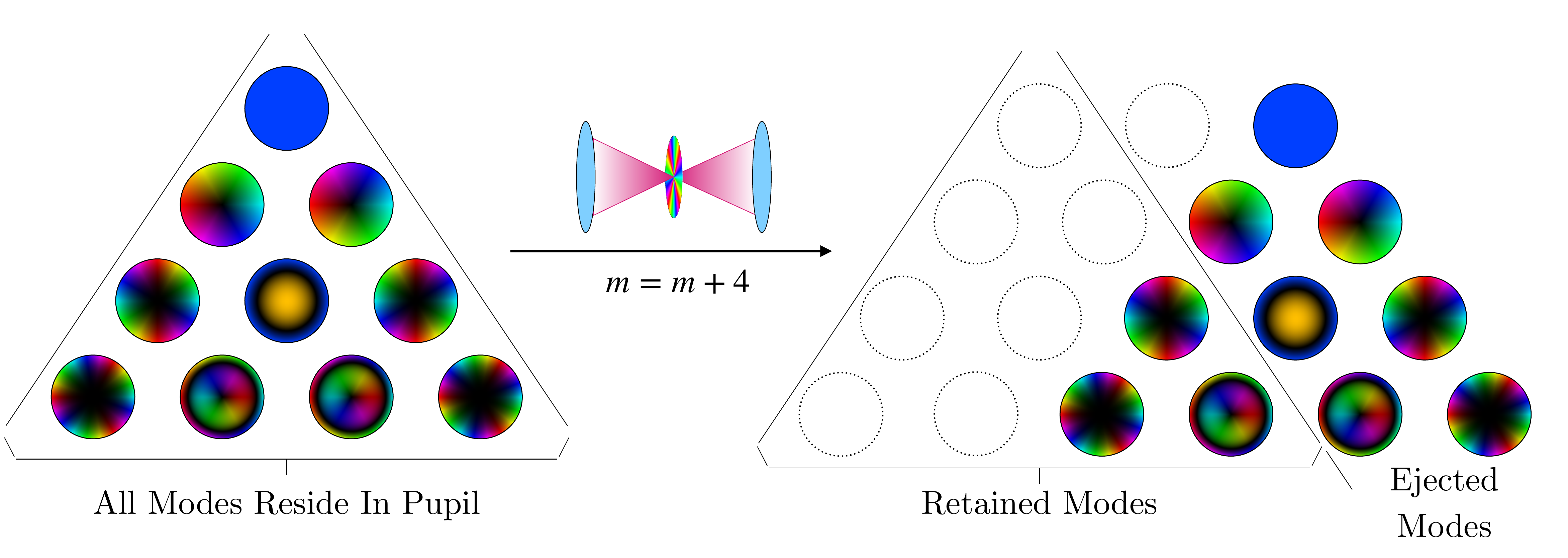}
    \caption{A mnemonic for understanding the action of a VPF on the complex Zernike pyramid, shown here for charge 4. We can imagine a VPF as shifting the entire pyramid left or right (positive or negative charge), moving some modes outside the pyramid boundary (oblique lines). Zernikes that cross these boundaries are ejected from the pupil. All other modes are retained, but transform into new Zernike polynomials $Z_{n'}^{m'} = Z_{n}^{m+l}$, where here $l=4$ but could in general be any even integer.}
    \label{fig:pyramid-cartoon}
\end{figure}

We can therefore understand a VPF as a beamsplitter, sorting modes by their relative OAM. Fig. \ref{fig:pyramid-cartoon} offers a simple mnemonic for visualizing this. Let us tile the Zernike modes into a pyramid-like structure as done in Fig. \ref{fig:zernikes} and then bound the entire structure by two lines that extend infinitely downward and intersect just above the piston mode. A VPF\textsubscript{$l$} then takes these modes and shifts them left and right, depending on the sign of $l$. Post-shift, some polynomials now reside outside the boundary, a mathematical event interpreted as complete destructive interference over the pupil. In contrast, modes whose shifted indices still correspond to valid Zernike polynomials are transformed into those modes. That is, they remain confined to the aperture's support. Hence, VPF\textsubscript{$l$} is a group beamsplitter, spatially separating the rightmost (or leftmost) $l/2$ pyramid diagonals from their complement.

\begin{figure}
    \centering
    \includegraphics[width=\linewidth]{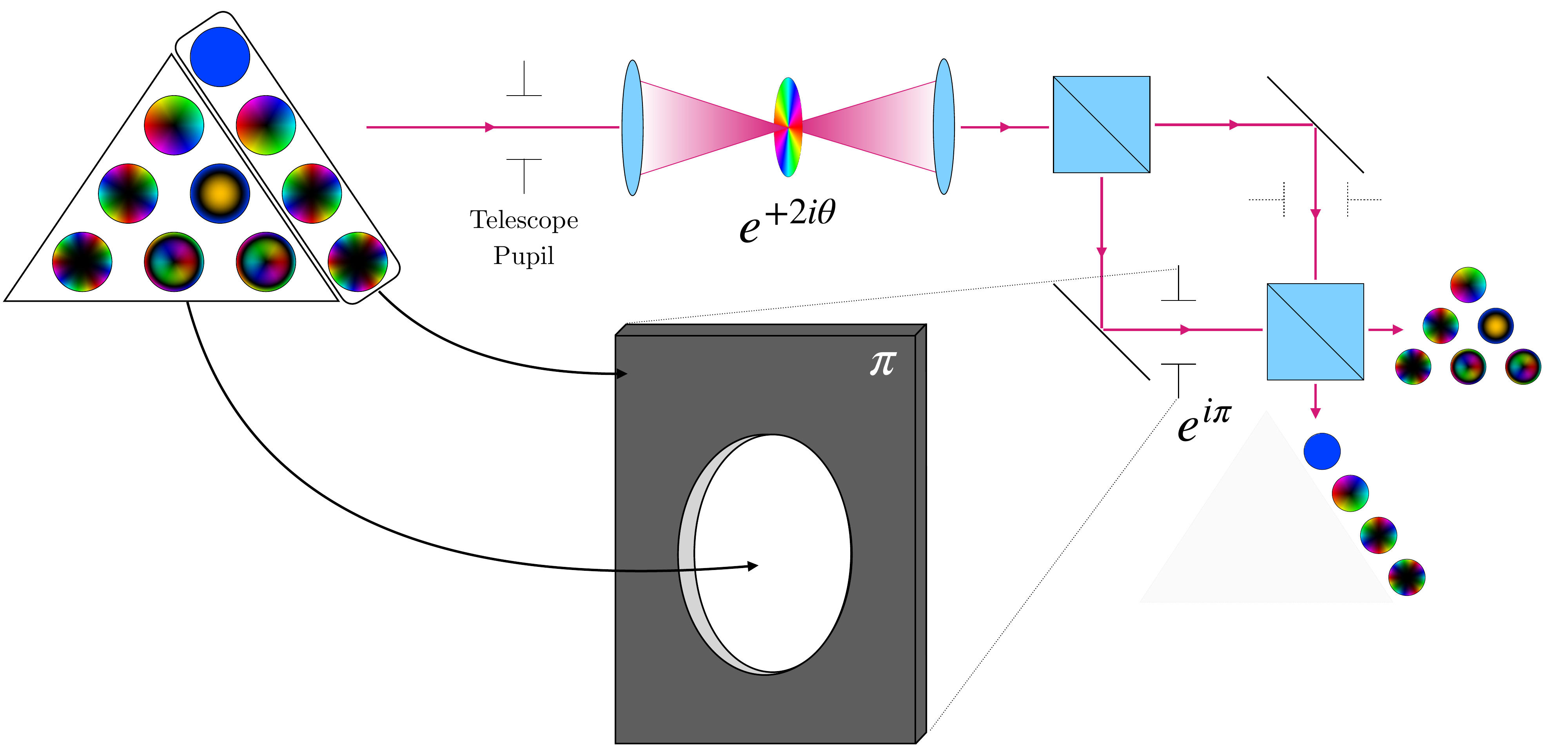}
    \caption{Our two-stage approach for separating pyramid edges in the complex Zernike pyramid. We first pass light through a VPF\textsubscript{2}, separating out the $m=n$ OAM modes from $m < n$. Following the second lens, we intercept the reforming pupil light with a balanced MZI, using phase shifting stops to bind the two different mode groups into orthogonal output ports.}
    \label{fig:first-stage}
\end{figure}

Separating mode groups in this manner now provides a constructive way to sort or filter \emph{all} the Zernike polynomials. We will present but one method (Fig. \ref{fig:first-stage}), but alternatives do exist. We first pass incident light through a VPF\textsubscript{2}. This sieves the Zernike pyramid's right edge from its orthogonal complement. However, before the pupil reforms we intercept the light with a balanced Mach-Zehnder interferometer (MZI), forming a copy of the pupil in each arm. Along one path we place a $\pi$ phase shifting stop, i.e., an iris that phase shifts the ejected modes by one-half waves. Modes following the second path are left to propagate. At the second beamsplitter, our asymmetric phase shifting forces the two mode groups to bind to different output arms. We now have independent control over both sets.

\begin{figure}
    \centering
    \includegraphics[width=\linewidth]{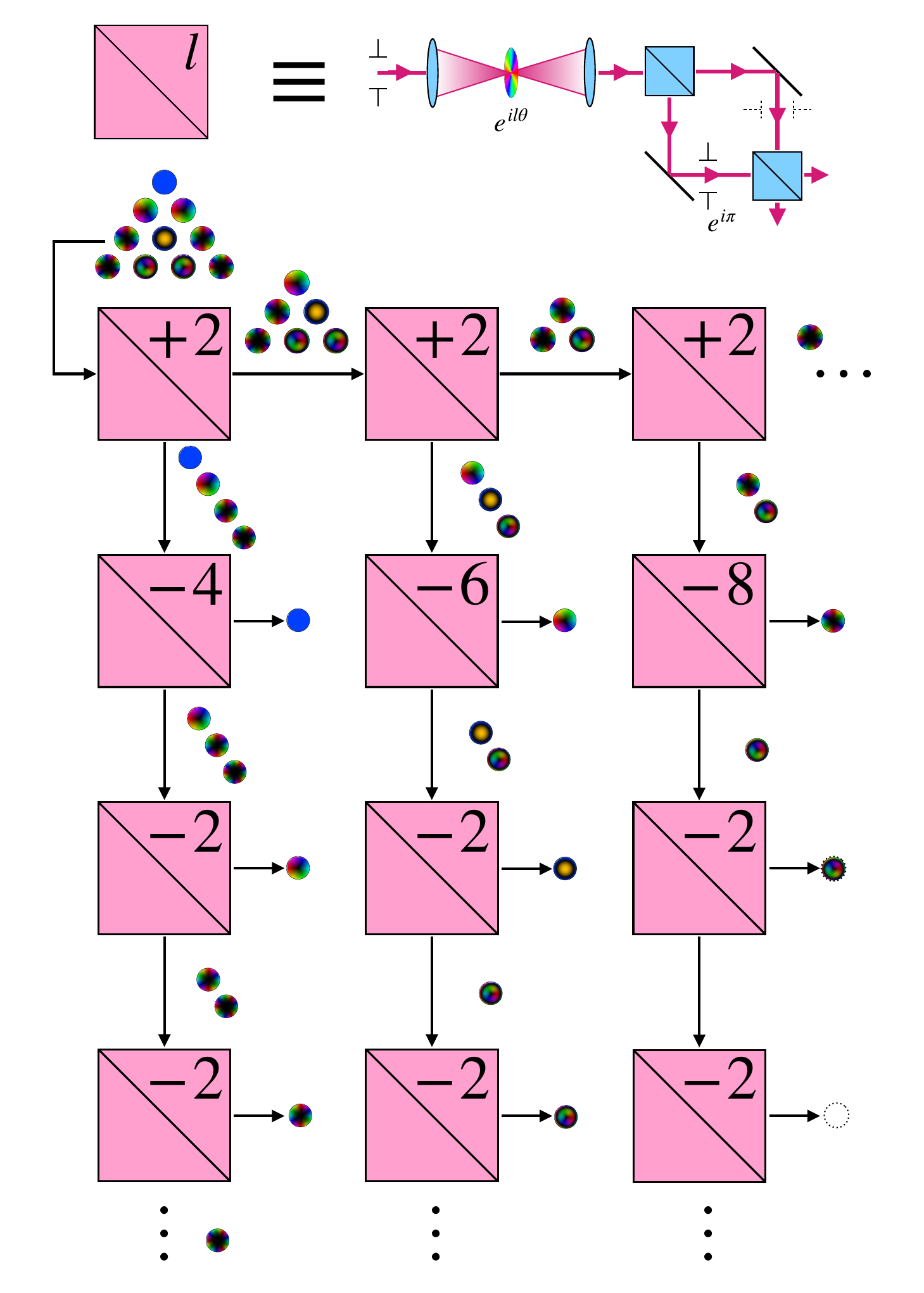}
    \caption{An optical processor effecting perfect complex Zernike decompositions. We rely on iterated use of the VPF-MZI developed in Fig. \ref{fig:first-stage} to sort Zernike pyramid diagonal-by-diagonal, and then one-by-one along each diagonal. Mode appearing immediately right and below a VPF-MZI optic represent the two groups just separated. When only a single mode appears we have perfectly isolated it.}
    \label{fig:recursive}
\end{figure}

This newfound control offers a recursive procedure for sorting all the complex Zernike modes. We schematize a general approach in Fig. \ref{fig:recursive}, which can be envisioned as an array of the VPF-MZI assemblies shown in Fig. \ref{fig:first-stage}. Each column is identical to within the charge of their second VPF-MZI group. Essentially, we eject diagonal mode groups one at a time using the VPF-MZIs in the top row. We will linearly index these columns by $k \geq 1$. Moving down a row, we now add a VPF-MZI having charge $-2(k+1)$. This undoes the prior $2k$ OAM shift introduced by the top row and then subtracts an additional two charge units, ejecting that diagonal's topmost mode whilst returning higher-modes to the pupil. Applying further charge -2 VPF-MZIs will then sieve the follow-on modes one-by-one, as the topmost remaining mode always rests against the pyramid's left edge. Theoretically, we can continue this process indefinitely, accessing as many modes as desired.

Before considering some illustrative use cases, we note that our protocol can also sort real-valued Zernikes. Unlike complex sorting, we will first pre-filter the field by introducing an inversion interferometer at the beginning of the optical train. For example, we can use a MZI with a single mirror in one path and two mirrors in the other path. Inspecting the output beams will show that $m<0$ modes are coupled into one output arm while $m\geq0$ modes are coupled into the other. In both arms we execute vortex filtering as before, separating the field into its constituent complex Zernikes. For a real-valued mode $(n,m)$, we then just collect light from the $(n,\pm|m|$) OAM modes, as these must have originated from the $\cos(m\phi)$ ($\sin(m\phi)$) functions if $m \geq 0$ $(m < 0)$.  

\section{Application to Wavefront Sensing and Coronagraphy}\label{sec:applications}
\begin{figure*}
    \centering
    \includegraphics[width=\linewidth]{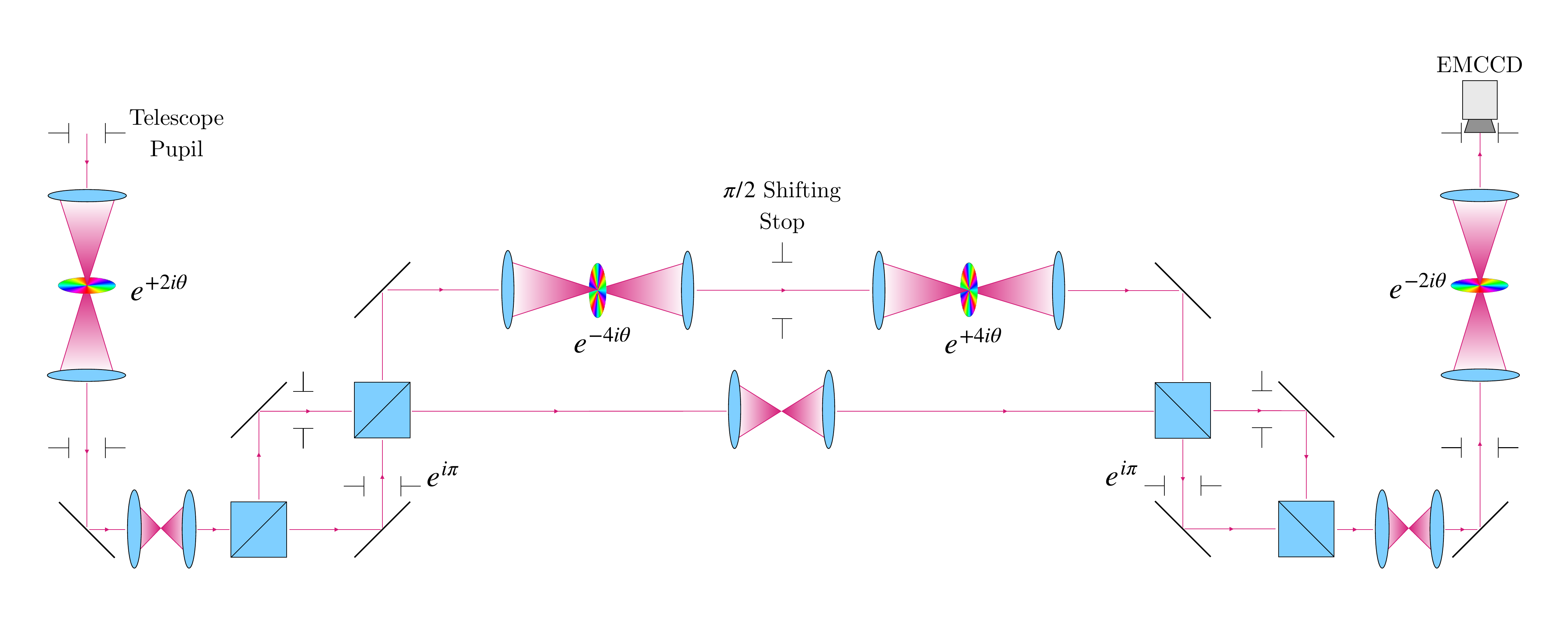}
    \caption{Hardware prescription for an ideal wavefront sensor. Collimated light (lines) from a guide star enters the pupil and is then filtered with a VPF\textsubscript{2}, separating the right pyramid edge from other modes. Following a potential second relay, an MZI with one $\pi$ phase shifting stop couples the two mode sets into orthogonal arms. Modes still residing in the pupil are merely relayed to the inversion optics, which eventually reconstitute the pupil. Rejected modes are sent into a VPF\textsubscript{-4}, which returns all but piston to the pupil. A $\pi/2$ phase shifting stop now operates exclusively on piston. Remaining optics invert the mode separation and recombine everything back in the pupil plane, finally effecting a quantum-optimal wavefront sensor. Note that by swapping the $\pi/2$-shifting mask for an opaque mask we generate a quantum-optimal coronagraph. In either case we then image the field with a photon-counting detector, shown here as an EMCCD.}
    \label{fig:paws/pac}
\end{figure*}

Prior work has shown that optimal WFS and coronagraphs implement nearly equivalent optical processing, differing only in their treatment of piston. Indeed, \cite{paterson_towards_2008, guyon_limits_2005} showed that an optimal wavefront sensor phase shifts piston by $\pi/2$ while \cite{guyon_theoretical_2006} showed that an optimal coronagraph eliminates piston altogether. Ref. \cite{chambouleyron_coronagraph-based_2024} recently made this connection explicit using the language of linear operators. Nevertheless, surgical operation on the piston mode, while leaving other modes untouched, remains an outstanding experimental challenge. Several Fourier-filtering sensors approach these limits \cite{bloemhof2004phase, 2023aoel.confE..79H, guyon_high_2010} with two such devices -- the bvWFS and the VC -- using a single VPF. But as we have shown, a VPF\textsubscript{2} ejects too many modes as \emph{all} $m=n$ polynomials are lost to the stop.

Adding a VPF\textsubscript{-4} to the bvWFS and VC optical trains addresses this problem perfectly (Fig. \ref{fig:vpf}). We first propagate through a VPF\textsubscript{2}, separating the Zernikes into two groups, one for which $m<n$ and the other for which $m=n$. Unlike a bvWFS or a VC, no processing occurs at the reimaged stop. We now relay the pupil into an MZI as just discussed split the modes into two sets: those having $m>n$ and $m\leq n$. Piston and all other modes on the Zernike pyramid's right edge are coupled out the top port and sent into a VPF\textsubscript{-4}, moving $m > 0$ modes back into the pupil while retaining piston's exclusion. A $\pi/2$ phase shift is now be applied to piston only, effecting an ideal WFS, or a standard Lyot stop can be applied, totally eliminating starlight and instituting a perfect coronagraph. Our remaining optics serve to invert the mode separation and reconstitute the pupil.

To our knowledge these optical layouts effectuate the most sensitive wavefront sensor and coronagraph ever proposed. While another quantum-optimal coronagraph did appear in \cite{deshler_experimental_2025}, that design sorted only four modes, limiting performance at large star-planet separations, while also being restricted to a single polarization and narrowband light. In contrast, the combination of VPF\textsubscript{$\pm$2} and VPF\textsubscript{$\pm$4} assemblies can be rendered achromatic and polarization-insensitive \cite{mawet_annular_2005, Swartzlander2006}. Phase shifters can be achromatized via the Pancharatnam-Berry effect \cite{Berry01111987, hariharan1994achromatic, bloemhof2014application} and mirrors can replace lenses. Our optics also process every mode simultaneously, making our procedure optimal for all aberrations in wavefront sensing and for unresolved coronagraphy at all star-planet separations.

\begin{figure*}
    \centering
    \includegraphics[width=\linewidth]{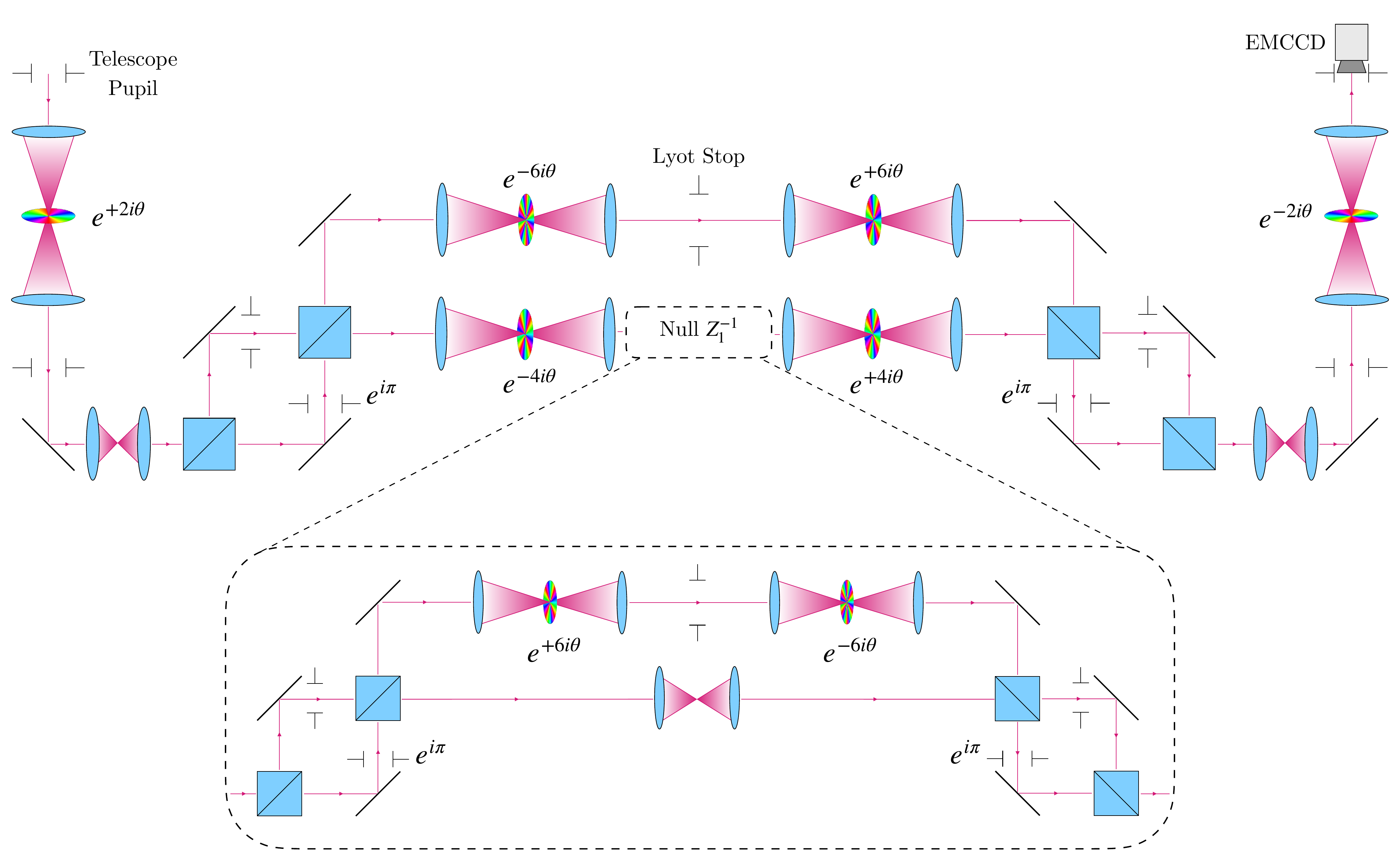}
    \caption{Optical layout for a perfect fourth-order coronagraph. Piston and tip are removed by swapping VPF$_{\pm4}$ in the perfect second-order coronagraph for VPF$_{\pm6}$. Zernike tip is nulled using an equivalent second-order system in the other arm. However, this additional subsystem now employs VPF\textsubscript{$\pm$6} in order to cross the Zernike pyramid's left edge over itself to eject tip.}
    \label{fig:4th-order}
\end{figure*}

We can also use VPFs to realize higher-order coronagraphs, as discussed by Belikov \textit{et. al.} \cite{belikov2021}. Such designs are critical for real observations where finite stellar sizes excite more than than just piston. The simplest of these is called a \textit{fourth-order coronagraph}; it eliminates the first three Zernike modes (piston, tip, and tilt) rather than just $Z_0^0$. This sacrifices planet throughput but improves robustness to telescope pointing error and stellar size. We can easily adapt our perfect coronagraph to function at fourth-order (Fig. \ref{fig:4th-order}). This is done by exchanging VPF${}_{\pm4}$ for VPF${}_{\pm6}$ in the perfect second-order coronagraph and swapping the relay in the orthogonal arm for another second-order coronagraph, now tailored to eliminate tip. The first of these modifications serves to null piston and tilt. The second treats tip as a piston mode with respect to its position in the truncated Zernike tower, using a similar combination of pyramid shifting to isolate it outside the pupil (recall Fig. \ref{fig:recursive}). Higher-order coronagraphs can be realized in a similar fashion by again increasing the VPF charges and adding an additional second order coronagraph where the relay resides.

\subsection{Extension to Arbitrary Apertures}\label{sec:arbitrary-apertures}As developed, our formalism only works with circularly symmetric imaging systems. Because real telescopes have spiders supporting the secondary mirror, the perfect coronagraphs and ideal WFS just presented are manifestly suboptimal. Achieving quantum limits for arbitrary pupils requires that we access the piston, tip, tilt, etc. modes native to the telescope's aperture. This can be accomplished by using a single-mode converter to reshape the aperture's native piston mode into that of Zernike piston. When done with high fidelity, the transformed high-order native modes are automatically orthogonal to Zernike piston. We can then employ our second-order coronagraph for circularly symmetric optics to sieve and finally null the fundamental mode. Application to wavefront sensing is similar. 

This subsystem could implemented using a few-plane MPLC or some other computer generated hologram. The reader may wonder why not exercise MPLC from the start? This is because MPLC-based mode sorting requires that we explicitly control the mapping between \emph{every} input and output mode. Here, we are only concerned with converting native piston into Zernike piston. While non-piston modes are uncontrolled, generally having their energy distributed both inside and outside of the aperture, this is irrelevant to piston sorting. We already know how to sunder light inside and outside of the pupil using an MZI. After the interferometer we can sort Zernike piston as before, since any remaining light from high-order native modes must reside in span$\{Z_0^0\}^\perp$. Interestingly, higher-order aperture-native modes can also be sieved in this way. Our work therefore provides a constructive procedure for generating arbitrary single-photon unitary transformations in the spirit of \cite{reck_experimental_1994}.

\section{Conclusion}\label{sec:conclusion}
We have presented a spatial mode sorting scheme based on vortex phase filtering for broadband, polarization-insensitive optical processing on finitely supported optical fields. Our method relies on a remarkable algebraic feature of the Zernike polynomials that permits selective expulsion of modes from the aperture. This feature, together with a possible suite of single-mode converters, provides a constructive procedure for demultiplexing light into any desired spatial mode basis. Our technique is in principle lossless, incurs zero crosstalk, and relies on hardware that is readily accessible. These attributes make VPF-based mode sorting ideal for near-term quantum metrology tasks, as was demonstrated using wavefront sensing and coronagraphy as case studies. 

However, several significant challenges still remain. For instance, using our approach for broadband mode sorting requires that the mode converters be achromatic. This is difficult to achieve in practice due to spectral dependence in Rayleigh-Sommerfeld propagation. An MPLC with a large number of planes partially address this problem, but its cost and training complexity quickly rise, so too its loss. PLs and PICs are undesirable for coupling reasons. Solving this problem will be central to future work.

Our proposal also requires a large number of MZIs for separating retained and ejected mode groups. These interferometers should be aligned to sub-wavelength precision, both within the MZI and relative to all VPFs. This is a formidable alignment challenge. For wavefront sensing and unresolved coronagraphy there are merely two MZIs, a difficult alignment exercise but far from impossible. However, for resolved coronagraphy, rejecting an additional Zernike row introduces another perfect coronagraph and two new VPFs. Thus, while vortex phase filtering offers a theoretically perfect mode sorting platform, determining its practical utility requires detailed tolerancing analyses and experimental validation.

\begin{backmatter}
%\bmsection{Acknowledgment}
%Additional information crediting individuals who contributed to the work being reported, clarifying who received funding from a particular source, or other information that does not fit the criteria for the funding block may also be included; for example, ``K. Flockhart thanks the National Science Foundation for help identifying collaborators for this work.'' 

\bmsection{Disclosures}
The authors declare no conflicts of interest.

%\bmsection{Data availability} Data underlying the results presented in this paper are not publicly available at this time but may be obtained from the authors upon reasonable request.

\bmsection{Data availability} No data were generated or analyzed in the presented research.

%\bmsection{Supplemental document}See Supplement 1 for supporting content.
\end{backmatter}

%%%%%%%%%%%%%%%%%%%%%%% References %%%%%%%%%%%%%%%%%%%%%%%%%
\bibliography{ref}

%%%%%%%%%%%%%%%%%%%%%%% Appendices %%%%%%%%%%%%%%%%%%%%%%%%%%
\appendix
\section{Propagating Zernike Polynomials Through An Even-Charge Vortex Phase Filter}\label{app:sec:vortex-math}
In this section we will produce a closed-form expression for complex-valued Zernike polynomials $Z_n^m$ after propagated through a VPF. As shown in Fig. \ref{fig:vpf}, we first bring the pupil-plane field, and thus each Zernike mode, to focus. Take note that we have assumed paraxial waves. According to Noll \cite{noll_zernike_1976} we can then write these \textit{Fourier-Zernike} modes as \begin{equation}
    \Tilde{Z}_n^m(\xi, \theta) = (-1)^{(n-m)/2}i^m\sqrt{(n+1)(2 - \delta_{m0})}\frac{J_{n+1}(\pi \xi/\lambda f)}{\pi \xi}e^{im\theta},
\end{equation} where $J_n$ is a Bessel function of the first kind, $\lambda$ is the light's wavelength, $D$ is the aperture diameter, and $f$ is the lens' focal length. We next apply the vortex phase $e^{il\theta}$. This shifts the shift the modes' OAM by l. We finally invert our imaging with another lens (ignoring possible magnification) to produce the final mode
\begin{equation}
    \Check{Z}_n^m(r, \phi) = c_{nm}\int_{0}^{\infty}d\xi \xi\frac{J_{n+1}(\pi D\xi/\lambda f)}{\xi}\int_{0}^{2\pi}d\theta e^{2\pi i\xi r \cos(\theta-\phi)}e^{i(m+l)\theta}, 
\end{equation}where $c_{nm}=(-1)^{(n-m)/2}i^m\sqrt{(n+1)(2 - \delta_{m0})}/\pi$ and the first exponential is the usual Fourier kernel expressed in polar coordinates.

Incredibly, this integral can be evaluated in closed-form for any even-order vortex phase shift. Upon substituting $\theta = \gamma + \phi$, we can identify the angular integral as an $(m+l)$-th order Bessel function \cite[\href{https://dlmf.nist.gov/10.9.E2}{(10.9.2)}]{NIST:DLMF} with attached vortex phase, viz., \begin{equation}
    \Check{Z}_n^m(r, \phi) = c_{nm}\int_{0}^{\infty}d\xi J_{n+1}(\pi D\xi/\lambda f)J_{m+l}(2\pi \xi r / \lambda f) e^{i(m+l)\phi}.
\end{equation} We now recall that $l$ is even by hypothesis and infer that $m+l$ has the same parity as $n$. Hence, for $|m + l| \leq n$ the radial integral reduces to a radial Zernike polynomial \cite{noll_zernike_1976}. Whenever $|m + l| > n$ our calculation is facilitated by the remarkable identity \cite[\href{https://dlmf.nist.gov/10.22.E64}{(10.22.64)}]{NIST:DLMF}\begin{multline}
    \label{app:eq:remarkable-identitty}
    \int_{0}^{\infty}d\xi J_{n+1}(\pi D\xi / \lambda f) J_{m + l}(2\pi \xi r / \lambda f) = \\\lambda f \begin{cases}
        0, & 2r < D\\
        \frac{(-1)^k}{4r}, & 2r = D\\
        \frac{D^{n + 1}\Gamma(n + k + 2)}{(2r)^{n + 2}k!}{}_2 F_1\left[-k, n + k + 2; n + 2; \left(\frac{D}{2r}\right)^2\right], & 2r > D
    \end{cases},
\end{multline}with ${}_2F_1$ denoting the hypergeometric function, $\Gamma$ being Euler's gamma function, and \begin{equation}
    k = (m + l - n - 2) / 2.
\end{equation} Equation \eqref{app:eq:remarkable-identitty} states that modes shifted outside the Zernike pyramid's support are null over the entire pupil, exactly as posited.
\end{document}